\documentclass[journal]{IEEEtran}

\usepackage{cite,graphicx,amssymb,amsmath,color,textcomp}
\newtheorem{remark}{\underline{Remark}}

\begin{document}

\title{  Transceiver Design for Ambient Backscatter Communication over Frequency-Selective Channels  }  

\author{Chong Zhang$^\dag$, Yao Qin$^\ddag$,  Gongpu Wang$^\ddag$, Ruisi He$^\ddag$, and Rongfei Fan$^\S$ \\
$^\dag$School of Electronic and Information Engineering, Beijing Jiaotong University, Beijing, China \\
$^\ddag$School of Computer and Information Technology, Beijing Jiaotong University, Beijing, China\\
$^\S$School of Information and Electronics, Beijing Institute of Technology, Beijing, China\\
Email: \{15212111, 15281087, gpwang\}@bjtu.edu.cn, ruisi.he@ieee.org,  fanrongfei@bit.edu.cn\\ }
\maketitle

\begin{abstract}
Existing studies about ambient backscatter communication
mostly assume flat-fading channels.
However,     frequency-selective channels  widely exist
 in  many   practical scenarios.
Therefore, this paper investigates  ambient backscatter communication systems
over frequency-selective channels.
In particular, we propose an interference-free transceiver design
to   facilitate    signal detection    at  the reader.
Our  design    utilizes
the cyclic prefix (CP)
of orthogonal frequency-division multiplexing (OFDM)  source
symbols, which can cancel the signal interference
and thus   enhance  the detection accuracy  at the reader.
Meanwhile, our design  leads to  no interference  on  the
existing OFDM communication systems.
Next  we suggest a chi-square based detector for
the reader and derive the optimal detection threshold.
Simulations are then  provided to corroborate our  proposed studies.
\end{abstract}


\section{Introduction}
Ambient   backscatter   \cite{David_2013SigComm},
 a  newborn   green   technology   for   the Internet of Things (IoT),
has attracted much attention from both academia and industry \cite{Wang_2016ITComm, Qian_2017_TWC, Yang_2018_TWC, Ma_2018_ICL, Ma_2015_multi_antenna, LiDong_Access}.
Ambient   backscatter      utilizes the
ambient radio frequency signals to enable the   backscatter   communications
of  low data-rate  devices such as tags or sensors,
  and can  free them from batteries.

\begin{figure}[t]
\centering
\includegraphics[height=65mm,width=85mm]{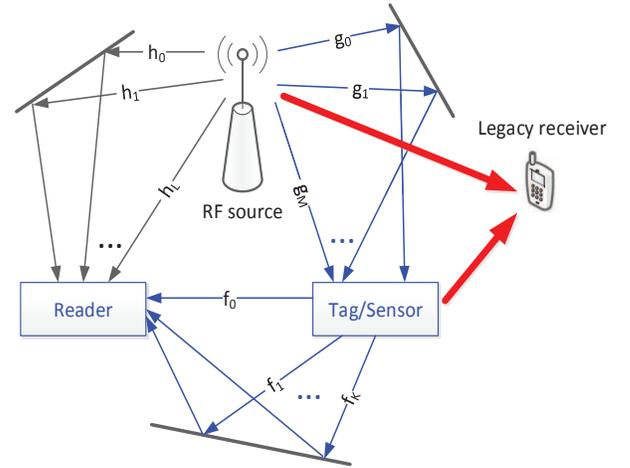}
\caption{System model. }
\label{fig:System_Model} \vspace{-5mm}
\end{figure}

A typical ambient   backscatter communication system
includes three components:  a radio frequency (RF) source, a tag (or a sensor),  and a reader,
as shown in Fig. \ref{fig:System_Model}.
The communication  process   between the tag and the reader  mainly contains two steps:
first, the tag  harvests energy from the  signals  of  the RF  source;
second, the tag  modulates its binary information
  onto the received RF signals and  then backscatters them  to   the reader.

Almost all existing studies  \cite{David_2013SigComm, Wang_2016ITComm, Qian_2017_TWC, Yang_2018_TWC, Ma_2018_ICL, Ma_2015_multi_antenna, LiDong_Access}
about ambient backscatter communication
are based on the assumption of   flat-fading channels.
However,    the frequency-selective channels  often exist in
 many  practical scenarios.
For   ambient backscatter communication systems,
  frequency-selective channels
  may  result in multiple copies of backscattered  signals at the reader,
   together with   multiple   source signals.
  Accordingly,  it is   one   challenging  problem  for  the reader
  to decode and  recover the tag signals.

In this paper, we investigate the
ambient backscatter communication systems
over frequency-selective channels  and
  propose a       transceiver design
to cope with  the signal detection challenge  at the reader.
Our  design    smartly        utilizes
the cyclic prefix (CP)
of orthogonal frequency-division multiplexing (OFDM) source
symbols,    which can   facilitate signal detection at the reader
via cancelling  the signal interference.
 Moreover,   different    from the transceiver design in \cite{Yang_2018_TWC},
our design leads to no
   interference  to the legacy receivers.
A chi-square based detector
 is then  proposed  and  the corresponding    optimal  detection threshold
   is derived.
Simulation results   show that
our  transceiver  design for the frequency-selective channels
is efficient    and
     achieves       low   bit error rate (BER)      due to  interference cancellation.

The rest of this paper is organized as follows:
Section \ref{sec:System Model} formulates the model of the ambient backscatter communication
system over frequency-selective channels.
Section \ref{sec:Transceiver Design}
proposes the transceiver design
and Section \ref{sec:Signal Detection at the Reader} derives the chi-square based detector
together with the optimal detection threshold.
Section \ref{sec:Simulation Results} provides
the simulation results
and finally Section \ref{sec:Conclusion} summarizes this paper.

\section{System Model}
\label{sec:System Model}
Consider an ambient backscatter communication system
 over frequency-selective channels
   in Fig. \ref{fig:System_Model}.
 The   multi-path    channels between the
  RF source and reader,
  the  RF source and
    tag,
   the reader and  tag
  are denoted by
  $h_l \,\,(l=0, 1, \cdots, L)$, $g_m \,\,(m=0, 1, \cdots, M)$, $f_k \,\,(k=0, 1, \cdots, K)$,
  respectively.
 Both the reader and the legacy receiver receive signals from the RF source and the tag
  over frequency-selective channels.

Suppose the   signal transmitted by the RF source is $s(n)$
with the zero-mean and the variance of $P_s$ and $s(n) \sim \mathcal{CN}(0,P_s)$.
Due to the multi-path channels $g(m)\,\, (m=0, 1, \cdots, M)$,  the signal
  arriving    at  the tag antenna
can be  given as
\begin{align}
x(n)=\sum_{m=0}^{M} g_m s(n-m).
\end{align}

The tag  next
modulates its own binary signal $B(n)$
onto the received signal $x(n)$
to   communicate with
 the reader via
backscattering  $x(n)$  or not.
Specifically,   the   tag   changes  its antenna impedance
   to   reflect    $x(n)$     to the reader
so as to indicate  $B(n)=1$;
and  when  indicating   $B(n)=0$,
  the tag switches   the impedance to a certain value
   so that no signal
 can be reflected.
 Assume that $B(n)=0$ and $B(n)=1$ are equiprobable.

Finally, the received signal at the reader can be expressed as
\begin{align}
y(n)=   \sum_{l=0}^{L} h_l s(n-l) + \eta \sum_{k=0}^{K} f_k B(n-k) x(n-k) +w(n),
\end{align}
where  $\eta$ represents the complex  attenuation inside the tag,
$w(n)$  denotes  the additive white Gaussian
noise (AWGN)  and we assume $w(n) \sim \mathcal{CN}(0,N_w)$.

\begin{remark}
The reader    aims   to  recover the tag signal $B(n)$
from the received signal $y(n)$.
 Nevertheless, since the binary signal $B(n)$   hides    in the received signal $y(n)$,
it is   inefficient to utilize the methods in traditional
point-to-point and relay communication systems
to realize  the  recovery of $B(n)$.
In addition, the frequency-selective channels   worsen this dilemma due to multiple copies
 of the source signals that appear in the received signals $y(n)$.
Consequently, a transceiver design together with
a signal detector are required to
achieve   accurate     recovery of $B(n)$,
which will  be introduced   in our later
 Section \ref{sec:Transceiver Design} and Section \ref{sec:Signal Detection at the Reader},
respectively.

\end{remark}

\section{Interference-free Transceiver Design}
\label{sec:Transceiver Design}
In this section, we describe an interference-free transceiver design,
whose implementation   mainly   consists of three crucial aspects:
the   tag  signal  design,
the signal interference cancelling    method,
and the discrete   Fourier transformation (DFT) operation.

\subsection{Tag  Signal  Design}

\begin{figure}[t]
\centering
\includegraphics[height=45mm,width=90mm]{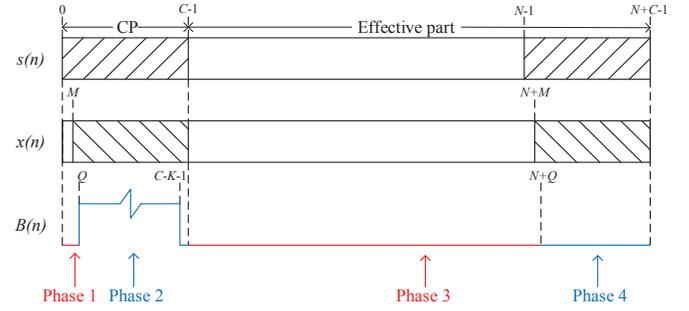}
\caption{The structures of the RF source signal $s(n)$, the tag signal $B(n)$, and the received signal $x(n)$  at the tag. }
\label{fig:Signal_Structure} \vspace{-3mm}
\end{figure}

With the assumption that the RF source emits OFDM symbols,
the structures of the RF source signal $s(n)$, the tag signal $B(n)$,
and the received signal $x(n)$ at the tag,
 are presented in Fig. \ref{fig:Signal_Structure}.\footnote{Since the OFDM technique is ubiquitous
 in current wireless systems such as
LTE and WiFi, it is reasonable to consider the RF source emitting OFDM symbols.}
We set $C$ and $N$ as the lengths of the CP and the effective part of the OFDM symbol, respectively.
The parameter $Q$ is defined as $Q={\rm max}\lbrace L, M, K\rbrace$.

Obviously, both $s(n)$ and $x(n)$
have repeating sequences, even if the signal $x(n)$ experiences the
multi-path channels $g(m)\,\, (m=0, 1, \cdots, M)$.
Besides, we divide one OFDM symbol period into four
phases for the designed tag signal $B(n)$.
In Phase 1, Phase 3, and Phase 4, no received signal $x(n)$  will be reflected, i.e., $B(n)=0$.
In this case,  the signals arriving at the reader directly come from the RF source.
However, in Phase 2,  the tag modulates its binary data onto the signal $x(n)$ from
 $n=Q$ to  $n=C-K-1$ while no signal is backscattered to the reader
in the rest of the Phase 2.
By exploiting the signals arriving at the reader in Phase 2 and Phase 4,
we can cancel the signal interference, which will be presented in
the next subsection.

\begin{remark}
It can be checked from Fig. \ref{fig:Signal_Structure} that the signal structure of $B(n)$
merely effects the samples in the CP of the OFDM symbol.
Since the CP will be removed at the legacy receiver,
this transceiver design at the tag will
lead to no interference to the legacy receivers.
\end{remark}

\subsection{Signal Interference Cancelling Method}
\label{subsec:Signal_interference_cancelling}
Denote the
received signals at the reader
in Phase 2 and Phase 4
as $y_{1}(n)\,\,\,(n=Q, \cdots, C-1)$ and $y_{2}(n)\,\,\,(n=N+Q, \cdots, N+C-1)$, respectively.
We can obtain
\begin{align}
y_{1}(n)=&    \sum_{l=0}^{L} h_l s(n-l) + \eta \sum_{k=0}^{K} f_k B(n-k) x(n-k) +w_1(n),  \label{eq:y1_received_signal}\\
y_{2}(n)=&    \sum_{l=0}^{L} h_l s(n-l) +w_2(n),  \label{eq:y2_received_signal}
\end{align}
where $w_1(n)$ and $w_2(n)$ are both AWGN. Assume that $w_1(n) \sim \mathcal{CN}(0,N_w)$ and
$w_2(n) \sim \mathcal{CN}(0,N_w)$.


Apparently, the term
$\sum_{l=0}^{L} h_l s(n-l)$ in (\ref{eq:y1_received_signal}) and (\ref{eq:y2_received_signal})
 carries no tag binary information
 and thus  is    the interference
for the tag signal recovery at the reader,
 which should    be  cancelled   so as to
 enhance     detection accuracy.

Signal interference cancelling is implemented via subtracting $y_2(n)$ from $y_1(n)$,
thus the received signals can be written as
\begin{align}
z(n)= & y_{1}(n+Q)-y_{2}(n+N+Q)\nonumber \\
    = & \eta \sum_{k=0}^{K} f_k B(n-k) x(n-k) +w_1(n) -w_2(n),\nonumber \\
    = & \eta \sum_{k=0}^{K} f_k B(n-k) x(n-k) +w_e(n),
\label{eq:Receive_signals_self_interference_elimitation}
\end{align}
where $n=0, 1, \cdots, C-Q-1$ and $w_e(n) \sim \mathcal{CN}(0,2N_w)$.

Assuming $T=C-Q-1$ and $R=C-Q-K-1$, we rewrite (\ref{eq:Receive_signals_self_interference_elimitation}) in matrix as

\begin{align}
\begin{pmatrix} z(0)            \\  z(1)   \\\vdots
\\  z(n)
\\\vdots
\\  z(T)
 \end{pmatrix}=\eta
 \begin{pmatrix} f_0 & 0 &  \cdots & 0 & \cdots & 0 & 0
\\  f_1 & f_0 &  \cdots & 0 & \cdots  & 0 & 0
\\  \vdots & \vdots &  \cdots & \vdots & \cdots & \vdots &\vdots
\\  f_K  & f_{K-1} &  \cdots & f_0 & \cdots & 0  & 0
\\  \vdots  & \vdots &  \cdots  & \vdots & \cdots & \vdots  & \vdots
\\ 0 & 0 &  \cdots & 0 & \cdots & f_0 & 0
\\ 0 & 0 &  \cdots & 0 & \cdots & f_1 & f_0
\\  \vdots & \vdots &  \cdots & \vdots & \cdots & \vdots &\vdots
\\ 0 & 0 &  \cdots & 0 & \cdots & f_{K-1} & f_{K-2}
\\ 0 & 0 &  \cdots & 0 & \cdots & f_K & f_{K-1}
\\ 0 & 0 &  \cdots & 0 & \cdots & 0 & f_K
 \end{pmatrix} \nonumber\\
 \begin{pmatrix} B(0)x(0)            \\  B(1)x(1)    \\\vdots
\\  B(n)x(n)
\\\vdots
\\  B(R)x(R)
 \end{pmatrix}+
 \begin{pmatrix} w_e(0)            \\  w_e(1)   \\\vdots
\\  w_e(n)
\\\vdots
\\  w_e(T)
 \end{pmatrix}.
\end{align}

\subsection{DFT Operation}
After signal interference cancelling at the reader,
let us construct the signal vector $\textbf{z}$ as
\begin{align}
\textbf{z}= & [z(0)+z(R+1), \cdots, z(T-R-1)+z(T),  \nonumber\\
&   \qquad\qquad\,\,\,\,\, z(T-R), \cdots, z(R-1), z(R)]^{\rm T}.
\end{align}

Define
\begin{align}
\textbf{b}= & [B(0), B(1), \cdots, B(n),  \cdots, B(R)],\\
\textbf{x}= & [x(0), x(1), \cdots, x(n),  \cdots, x(R)]^{\rm T},\\
\textbf{w}= & [w_e(0)+w_e(R+1), \cdots, w_e(T-R-1)+w_e(T), \nonumber\\
 & \qquad\qquad\;\, w_e(T-R), \cdots, w_e(R-1), w_e(R)]^{\rm T}.
\end{align}

Denote $\textbf{F}$ as the $(R+1)\times (R+1)$  DFT matrix
with the  $(p,q)$th element $\textbf{F}_{pq}=\exp (-j2\pi pq/(R+1))$.
Let us consider a Toeplitz matrix $\textbf{T}$, which possesses the first row of $\textbf{t}_{\textbf{r}}=[f_0, 0, \cdots, 0,  f_K, f_{K-1}, \cdots, f_k,  \cdots, f_1]$ and the first column of $\textbf{t}_{\textbf{c}}=[f_0, f_1, \cdots, f_k,  \cdots, f_K, 0, \cdots, 0]^{\rm T}$.

Consequently, we   reconstruct the signal vector $\textbf{z}$ based on DFT as, i.e., DFT outputs $\tilde{\textbf{z}}$
\begin{align}
\tilde{\textbf{z}}= & \,\textbf{F}\textbf{z}  \nonumber \\
= & \,\eta \textbf{F}  \textbf{T} \cdot {\rm diag}[\textbf{b}] \cdot \textbf{x} + \textbf{F}\textbf{w} \nonumber  \\
= & \,\eta \cdot {\rm diag}\left[\tilde{\textbf{f}}^{\rm T}\right] \cdot {\rm diag}[\textbf{b}] \cdot \tilde{\textbf{x}} + \tilde{\textbf{w}},
\end{align}
where
\begin{align}
\tilde{\textbf{z}}=\, & [\tilde{z}(0), \tilde{z}(1), \cdots, \tilde{z}(n),  \cdots, \tilde{z}(R)]^{\rm T}=\textbf{F}\textbf{z},\\
\tilde{\textbf{x}}=\, & [\tilde{x}(0), \tilde{x}(1), \cdots, \tilde{x}(n),  \cdots, \tilde{x}(R)]^{\rm T}=\textbf{F}\textbf{x},\\
\tilde{\textbf{w}}=\, & [\tilde{w}_e(0), \tilde{w}_e(1), \cdots, \tilde{w}_e(n),  \cdots, \tilde{w}_e(R)]^{\rm T}=\textbf{F}\textbf{w},\\
\tilde{\textbf{f}}=\, & [\tilde{f}_0, \tilde{f}_1, \cdots, \tilde{f}_n,  \cdots, \tilde{f}_R]^{\rm T}=\textbf{F}\textbf{t}_{\textbf{r}}^{\rm T}.
\end{align}

According to the central limit theorem (CLT) \cite{Papoulis_2002_Probability_book}, we assume $\tilde{x}(n)\sim \mathcal{CN}(0, P_x)$, $\tilde{w}_e(n)\sim \mathcal{CN}(0, P_w)$ and $\tilde{f}_n\sim \mathcal{CN}(0, P_f)$,
 where
\begin{align}
P_x=& (R+1)P_s\sum_{m=0}^{M}|g_m|^2,\\
P_w=& 2(T+1)N_w,\\
P_f=& \sum_{k=0}^{K}|f_k|^2.
\end{align}

\section{Chi-square Based Signal  Detection at the Reader}
\label{sec:Signal  Detection at the Reader}
In this section, the chi-square based detector
together with the optimal detection threshold
are derived via the maximum likelihood (ML) principle.
The   corresponding  BER expression
is   also  obtained to evaluate the
  detection   performance.

\subsection{Chi-square Based Detector}
Due to the lower data-rate of the  tag signal $B(n)$ than that of the signal $\tilde{z}(n)$, we suppose the signal $B(n)$ remains
equivalent within $W$ samples of $\tilde{z}(n)$.
Let us construct the test statistic for detecting $B(n)$ as
\begin{align}
\Gamma_t=\frac{1}{P_w}\sum_{n=(t-1)W+1}^{tW} \left|\tilde{z}(n)\right|^{2},
\end{align}
where $t=1, 2, \cdots, T$, and  $\tilde{z}(n)$ is expanded as
\begin{align}
\tilde{z}(n)= \left\{
\begin{aligned}
&\tilde{w}_e(n),     &     {\rm if}\,\,\,B(n)=0,\\
& \eta \tilde{f}_n\tilde{x}(n)+\tilde{w}_e(n),      &          {\rm if}\,\,\,B(n)=1.  \\
\end{aligned}
\right.
\end{align}

It can be readily checked that
\begin{align}
\Gamma_t= \left\{
\begin{aligned}
&M_t,     &     {\rm if}\,\,\,B(n)=0,\\
&J_t+M_t+V_t,      &          {\rm if}\,\,\,B(n)=1,  \\
\end{aligned}
\right.
\end{align}
where
\begin{align}
M_t= & \frac{1}{P_w}\sum_{n=(t-1)W+1}^{tW} \left|\tilde{w}_e(n)\right|^{2},\\
J_t= & \frac{1}{P_w}\sum_{n=(t-1)W+1}^{tW} (|\eta|^2 \left|\tilde{f}_n\right|^2\left|\tilde{x}(n)\right|^2),\\
V_t= & \frac{1}{P_w}\sum_{n=(t-1)W+1}^{tW} (2\mathcal{R}\lbrace\eta \tilde{f}_n\tilde{x}(n){\tilde{w}^{*}_e(n)}\rbrace).
\end{align}

Let $\mathcal{H}_1$ and $\mathcal{H}_0$ represent $B(n)=1$ and $B(n)=0$, respectively.
Apparently,
under $\mathcal{H}_1$, the test statistic $\Gamma_t$ follows
the noncentral chi-square distribution
with $W$ degrees of freedom and noncentrality parameter $\lambda=W\gamma$ \cite{Horgan_2013_approximation_chisquare},
i.e., $\Gamma_t\sim \chi^2_W(\lambda)$,
where $\gamma$ is the detection signal-to-noise ratio (SNR) that   can be calculated as
 \begin{align}
 \gamma=\frac{|\eta|^2P_{x}P_{f}}{P_w}=\frac{(R+1)|\eta|^2P_{s}\sum_{m=0}^{M}|g_m|^2\sum_{k=0}^{K}|f_k|^2}{2(T+1)N_w}.
 \end{align}
While under $\mathcal{H}_0$,
$\Gamma_t$ follows
the chi-square distribution
with $W$ degrees of freedom, which is denoted as $\Gamma_t\sim \chi^2_W(0)$.

Therefore, the probability density function (PDF)
of $\Gamma_t$ under $\mathcal{H}_0$
is obtained as
\begin{align}
{\rm Pr}(\Gamma_t|\mathcal{H}_0)=f_0(\Gamma_t, W),
\end{align}
where
\begin{align}
f_0(x, n)= \left\{
\begin{aligned}
&\frac{1}{2^{\frac{n}{2}}\Gamma(n/2)}{\rm e}^{-\frac{x}{2}}x^{\frac{n}{2}-1},     &     {\rm if}\,\,\,x>0,\\
&0,      &          {\rm if}\,\,\,x\leq 0,  \\
\end{aligned}
\right.
\label{eq:PDF_B(n)=0}
\end{align}
and $\Gamma(x)=\int_{0}^{\infty}{\rm e}^{-t}t^{x-1}{\rm d}t$ as the Gamma function \cite{function_book}.

Similarly, under $\mathcal{H}_1$, the PDF
of $\Gamma_t$ is
\begin{align}
{\rm Pr}(\Gamma_t|\mathcal{H}_1)=f_1(\Gamma_t, W, \lambda),
\end{align}
where
\begin{align}
f_1(x, n, \lambda)= \left\{
\begin{aligned}
&\frac{1}{2}\left(\frac{x}{\lambda}\right)^{\frac{n-2}{4}}{\rm e}^{-\frac{x+\lambda}{2}}I_{\frac{n}{2}-1}\left(\sqrt{\lambda x}\right),     &     {\rm if}\,\,\,x>0,\\
&0,      &          {\rm if}\,\,\,x\leq 0,  \\
\end{aligned}
\label{eq:PDF_B(n)=1}
\right.
\end{align}
and $I_{r}(u)$ is the $r$-order modified Bessel function of the first kind \cite{function_book}
\begin{align}
I_{r}(u)=\frac{\left(\frac{1}{2}u\right)^r}{\sqrt{\pi}\Gamma(r+\frac{1}{2})}\int_{0}^{\pi}{\rm e}^{u\cos \theta}\sin^{2r}{\theta}{\rm d}\theta.
\end{align}

\begin{figure}[t]
\centering
\includegraphics[height=72mm,width=95mm]{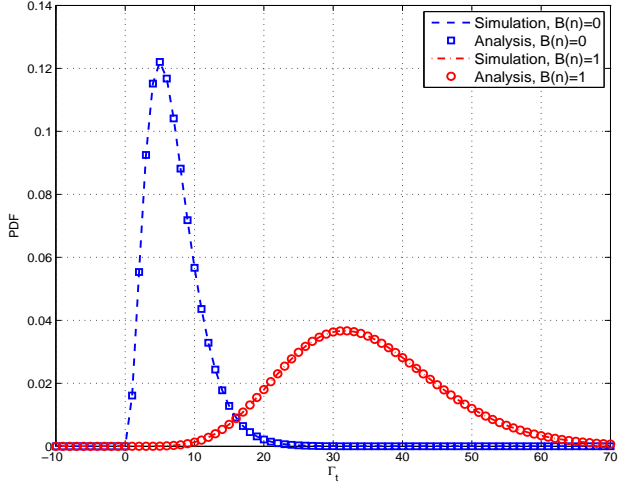}
\caption{An example for the PDFs of $\Gamma_t$ under two conditions $\mathcal{H}_0$ and $\mathcal{H}_1$. }
\label{fig:PDF} \vspace{-3mm}
\end{figure}

Consequently, the chi-square based detector can be made through the ML principle  as
\begin{align}
\hat{B}(n)=\mathop {\arg\max} \limits_{B(n)\in \lbrace 0,1\rbrace}{\rm Pr}(\Gamma_t|B(n)),
\label{eq:ML_detection_rule}
\end{align}
where ${\rm Pr}(\Gamma_t|B(n))$ is the probability density function (PDF) of $\Gamma_t$ given $B(n)$.
We can also reformulate the ML detection rule (\ref{eq:ML_detection_rule}) as
\begin{align}
\hat{B}(n)= \left\{
\begin{aligned}
&0,     &     {\rm if}\,\,\,{\rm Pr}(\Gamma_t|\mathcal{H}_0)>{\rm Pr}(\Gamma_t|\mathcal{H}_1),\\
&1,      &          {\rm if}\,\,\,{\rm Pr}(\Gamma_t|\mathcal{H}_0)<{\rm Pr}(\Gamma_t|\mathcal{H}_1).  \\
\end{aligned}
\right.
\end{align}

One example of the PDFs
of
$\Gamma_t$
under two conditions $\mathcal{H}_0$ and $\mathcal{H}_1$
is presented in Fig. \ref{fig:PDF}.

\subsection{Optimal Detection Threshold}
The optimal detection threshold $T_h$
of this ML detector
can be derived by setting that the PDF under $\mathcal{H}_0$ equals to that under $\mathcal{H}_1$
\begin{align}
{\rm Pr}(\Gamma_t|\mathcal{H}_0)={\rm Pr}(\Gamma_t|\mathcal{H}_1)\big|_{\Gamma_t=T_h}.
\label{eq:Optimal_threshold}
\end{align}

Define
\begin{align}
I=\int_{0}^{\pi}{\rm e}^{\sqrt{W\gamma T_h}}{\rm e}^{\cos \theta}\sin^{W-2}{\theta}{\rm d}\theta.
\end{align}
Substituting (\ref{eq:PDF_B(n)=0}) and (\ref{eq:PDF_B(n)=1})
into (\ref{eq:Optimal_threshold}) will produce
\begin{align}
\frac{T_h^{\frac{W-2}{4}}{\rm e}^{-\frac{T_h+W\gamma}{2}}\left(W\gamma T_h\right)^{\frac{W-2}{4}}}{2^{\frac{W}{2}}\sqrt{\pi}\left(W\gamma\right)^{\frac{W-2}{4}}\Gamma\left(\frac{W}{2}-\frac{1}{2}\right)}I
=\frac{{\rm e}^{-\frac{T_h}{2}}T_h^{\frac{W-2}{2}}}{2^{\frac{W}{2}}\Gamma{\left(\frac{W}{2}\right)}},
\end{align}
which can be further simplified as
\begin{align}
\frac{{\rm e}^{-\frac{W\gamma}{2}}I}{\sqrt{\pi}\Gamma{\left(\frac{W}{2}-\frac{1}{2}\right)}}=\frac{1}{\Gamma\left(\frac{W}{2}\right)}.
\label{eq:same_PDF_for_threshold}
\end{align}
After exerting some mathematical
manipulations in (\ref{eq:same_PDF_for_threshold}),
 one obtains
\begin{align}
T_h=\left(\ln{\frac{\sqrt{\pi}\Gamma{\left(\frac{W}{2}-\frac{1}{2}\right)}}
{{\rm e}^{-\frac{W\gamma}{2}}\Gamma\left(\frac{W}{2}\right)\int_{0}^{\pi}{\rm e}^{\cos \theta}\sin^{W-2}{\theta}{\rm d}\theta}}\right)^2\bigg/(W\gamma).
\end{align}

Therefore, the detection  rule can be summarized as
\begin{align}
\hat{B}(n)= \left\{
\begin{matrix}
0,  & {\rm if}\,\,\,  \Gamma_t<T_h,\\
1,   &{\rm if}\,\,\,  \Gamma_t>T_h.
\end{matrix}
\right.
\end{align}

\subsection{BER Performance}
Define $p_0={\rm Pr}(\hat{B}(n)=1|B(n)=0)$ and $p_1={\rm Pr}(\hat{B}(n)=0|B(n)=1)$
as the probability of false alarm and the probability of missing detection, separately.

The BER of the chi-square based detector is given by
\begin{align}
P_e= & {\rm Pr}(B(n)=0)p_0+{\rm Pr}(B(n)=1)p_1\nonumber\\
   = & \frac{1}{2}\left(p_0+p_1\right).
\end{align}
By utilizing the approximations in \cite{Horgan_2013_approximation_chisquare},
we can further derive the BER $P_e$
as
\begin{align}
P_e\approx\frac{1}{2}Q\left(\frac{T_h-W}{\sqrt{2W}}\right)+\frac{1}{2}Q\left(\frac{W(1+\gamma)-T_h}{\sqrt{2W(1+2\gamma)}}\right),
\label{eq:approximation_for_analysis_BER}
\end{align}
where $Q(\cdot)$ denotes the Gaussian Q-function   \cite{function_book}
 \begin{align}
Q(x)=\frac{1}{\sqrt{2\pi}}\int_{x}^{\infty}{\rm e}^{-\frac{t^2}{2}}{\rm d}t.
\end{align}

\section{Simulation Results}
\label{sec:Simulation Results}
In this section, numerical results are
provided to assess the  performance of  proposed chi-square based
 detector.
All the channels follow Gaussian distributions
with the zero-mean and unit-variance.
The  number of channel taps   $L+1$, $M+1$ and $K+1$ are assumed to  be $6$ in following simulations.
We set  the attenuation $|\eta|$, the noise power $N_w$ and
the length of CP $C$
   as $0.5$, $1$ and $256$, separately.
We also exert $10^7$
Monte Carlo trials
for every experiment
to examine BER performance of the chi-square based
 detector.

\begin{figure}[t]
\centering
\includegraphics[height=72mm,width=95mm]{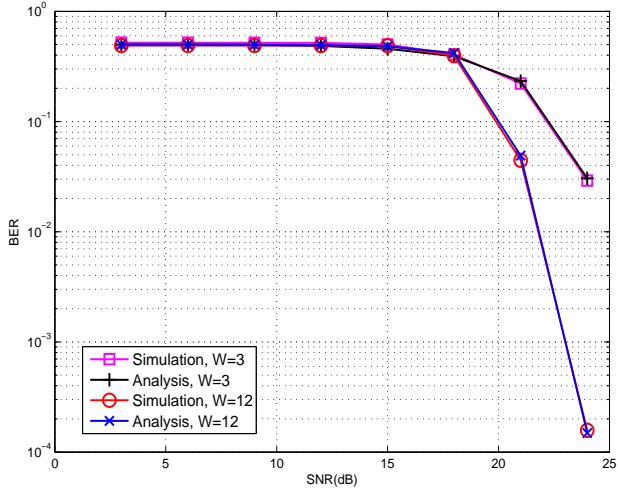}
\caption{BER for the chi-square based detector versus SNR. }
\label{fig:BER_SNR_simulation} \vspace{-3mm}
\end{figure}

Fig. \ref{fig:BER_SNR_simulation}
plots the BER curves versus SNR
for the chi-square based
 detector with the optimal detection threshold.
We set  the number of averaging samples $W$
as  $3$ and $12$, respectively.
As seen,
the BER performance could be enhanced with enlarging SNR or $W$.

\begin{figure}[t]
\centering
\includegraphics[height=72mm,width=95mm]{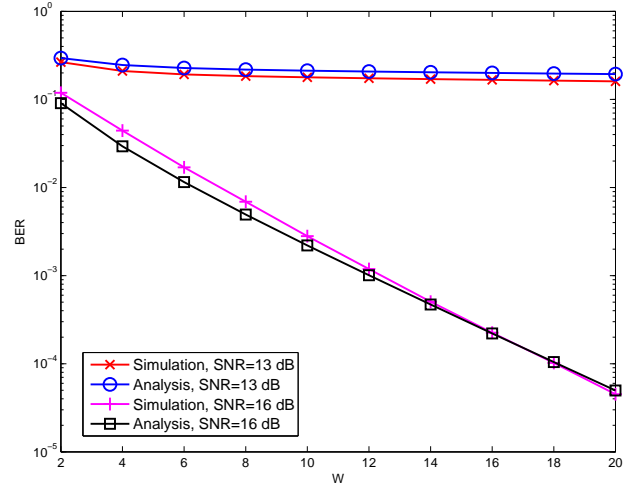}
\caption{BER for the chi-square based detector versus  the    averaging  number  W. }
\label{fig:BER_W_simulation} \vspace{-3mm}
\end{figure}

Fig. \ref{fig:BER_W_simulation}
depicts the BER curves versus the number of averaging samples $W$
with different SNR for our detector.
We choose
SNR as 13 dB and 16 dB, separately.
It is found that
 the BER performance is
improved with increasing $W$.
Besides,
due to the exploitation of the approximation (\ref{eq:approximation_for_analysis_BER})
for analytical BER,
there is a small gap between the simulation and analytical results
in Fig. \ref{fig:BER_W_simulation}.

\section{Conclusion}
\label{sec:Conclusion}
This paper focused on the data transmission of the ambient backscatter communication
systems over frequency-selective channels.
 A novel interference-free transceiver design, which  exploits
the CP structure of OFDM symbols to cancel the signal
interference,
was proposed to facilitate  interference  cancellation  and   signal detection  in such scenario.
Moreover, this transceiver design
led to no interference  to the legacy receivers
since the CP will be removed at the legacy receiver.
Furthermore, a chi-square based detector was derived,
as well as the optimal detection threshold.
Finally, the BER performance of the chi-square based detector was
evaluated via Monte Carlo simulations.  It was shown that the
proposed  transceiver design  is   efficient and the chi-square
based detector demonstrates  satisfying BER performance.

\end{document}